# Transform coder identification based on quantization footprints and lattice theory

Marco Tagliasacchi[1], Marco Visentini-Scarzanella[2], Pier Luigi Dragotti[2], Stefano Tubaro[1]

**Abstract**

Transform coding is routinely used for lossy compression of discrete sources with memory. The input signal is divided into $N$-dimensional vectors, which are transformed by means of a linear mapping. Then, transform coefficients are quantized and entropy coded. In this paper we consider the problem of identifying the transform matrix as well as the quantization step sizes. We study the challenging case in which the only available information is a set of $P$ transform decoded vectors. We formulate the problem in terms of finding the lattice with the largest determinant that contains all observed vectors. We propose an algorithm that is able to find the optimal solution and we formally study its convergence properties. Our analysis shows that it is possible to identify successfully both the transform and the quantization step sizes when $P \geq N + \delta$ where $\delta$ is a small integer, and the probability of failure decreases exponentially to zero as $P - N$ increases.

## I. INTRODUCTION

Transform coding has emerged over the years as the dominating compression strategy. Transform coding is adopted in virtually all multimedia compression standards including image compression standards such as JPEG [1] and JPEG 2000 [2], [3] and video compression standards such as, for example, H.264/AVC [4] and HEVC [5]. This is due to the fact that transform coders are very effective and yet computationally inexpensive since the encoding operation is divided into three relatively simple steps: the computation of a linear transformation of the data, scalar quantization of each coefficient, and entropy coding.

[1] Dipartimento di Elettronica e Informazione, Politecnico di Milano, P.zza Leonardo da Vinci, 32 20133 - Milano, Italy - E-mail: marco.tagliasacchi@polimi.it, stefano.tubaro@elet.polimi.it, [2] CSP Group, EEE Department, Imperial College London, Exhibition Road, London SW7-2AZ, United Kingdom, E-mail: p.dragotti@imperial.ac.uk, marcovs@imperial.ac.uk

Transform coding has been widely studied in the last decades and many important results and optimality conditions have been derived. For example, it is well known that, for Gaussian sources, the Karhunen-Loève Transform (KLT) is the optimal transform [6][7]. Moreover, the analysis of popular transform coders used in image compression has also led to new insights and new interesting connections between compression and non-linear approximation theory [8]. In particular, this analysis has also clarified why the wavelet transform is the best transform for compressing piecewise regular functions [9][10]. Further insights on the interplay between linear transform, quantization and entropy coding can be found in [11] for the case of integer-to-integer transforms.

Due to its centrality to any type of multimedia data, transform coding theory is now extensively used in a new range of applications that rely on the possibility of reverse-engineering complex chains of operators starting from the available output signals. Indeed, the lifespan of a multimedia signal is virtually unbounded. This is due to the ability of creating copies and the availability of inexpensive storage options. However, signals seldom remain identical to their original version. As they pass through processing chains, some operators, including transform coding, are bound to leave subtle characteristic footprints on the signals, which can be identified in order to uncover their processing history. This insight might be extremely useful in a wide range of scenarios in the field of multimedia signal processing at large including, e.g.,: i) forensics, in order to address tasks such as source device identification [12] or tampering detection [13][14]; ii) quality assessment, to enable no-reference methods that rely solely on the received signals [15][16]; iii) digital restoration, which requires prior knowledge about the operations that affected a digital signal [17].

In this context, several works have exploited the footprints left by transform coding. In [18], a method was proposed to infer the implementation-dependent quantization matrix template used in a JPEG-compressed image. Double JPEG compression introduces characteristic peaks in the histogram of DCT coefficients, which can be detected and used, e.g, for tampering localization [19][14]. More recently, similar techniques were applied to video signals for the cases of MPEG-2 [20][21], MPEG-4 [22][23] and H.264/AVC [24].

All the aforementioned works require prior knowledge of the type of standard being considered. This implies that the specific transform in use is assumed to be known, whereas the quantization step sizes need to be estimated. In practice, it might be useful to be able to infer which transform was used in order to understand, for example, whether an image was compressed using the DCT-based JPEG or the wavelet-based JPEG 2000 and, in the latter case, which wavelet transform was used. Similarly, it would be good to be able to infer if a video sequence was compressed using MPEG-2, MPEG-4 or H.264/AVC. Some efforts in this direction can be found in [25].

Most of the above methods focus only on a specific type of multimedia signal (e.g., only images or only videos) and are to some extent heuristic. It is therefore natural to try and develop a universal theory of transform coder identification that is independent of the specific application at hand. To this end, in this paper we consider a general model of transform coding that can be tailored to describe a large variety of practical implementations that are found in lossy coding systems, including those adopted in multimedia communication. Specifically, a 1-dimensional input signal is encoded by partitioning it into non-overlapping $N$-dimensional vectors, which are then transformed by means of a linear mapping. Then, transform coefficients are quantized and entropy coded. At the decoder, quantization symbols are entropy decoded and mapped to reconstruction levels. Then, the inverse transform is applied to obtain an approximation of the signal in its original domain.

Given the output produced by a specific transform coding chain, we investigate the problem of identifying its parameters. To this end, we assume both the size and the alignment of the transform to be known, as they can be estimated with methods available in the literature [21][18]. We propose an algorithm that receives as input a set of $P$ transform decoded vectors embedded in a $N$-dimensional vector space and produces as output an estimation of the transform adopted, as well as the quantization step sizes, whenever these can be unambiguously determined. We leverage the intrinsic discrete nature of the problem, by observing the fact that these vectors are bound to belong to a $N$-dimensional lattice. Hence, the problem is formulated in terms of finding a lattice that contains all observed vectors. We propose an algorithm that is able to solve the problem and we formally study its convergence properties. Our analysis shows that it is possible to successfully identify both the transform and the quantization step sizes with high probability when $P > N$. In the experiments we found that an excess of approximately 6-7 observed vectors beyond the dimension $N$ of the space is generally sufficient to ensure successful convergence. In addition, the complexity of the algorithm is shown to grow linearly with $N$.

It is important to mention that the method used to solve the problem addressed in this paper is related to Euclid's algorithm, which is used to find the greatest common divisor (GCD) in a set of integers. Indeed, when $N = 1$ and $P = 2$, the proposed method coincides with Euclid's algorithm. However, in this case the problem reduces to estimating the quantization step size, as the transform is trivially defined.

Note that, lattice theory has been widely used for source and channel coding (e.g., [26], [27], [28]). However, to the best of the authors' knowledge, this theory has not been employed to address the problem of identifying a linear mapping using the footprint left by quantization. Only [29] uses similar principles but their goal is to investigate the color compression history, i.e., the colorspace used in JPEG compression. Therefore, the solution proposed is tailored to work in a 3-dimensional vector space, thus avoiding the challenges that arise in higher dimensional spaces.

Also, it is important not to confuse the problem addressed in this paper with the classical problem of lattice reduction [28]. In the latter case, given a basis for a lattice, one seeks an equivalent basis matrix with favorable properties. Usually, such a basis consists of vectors that are short and with improved orthogonality. There are several definitions of lattice reduction with corresponding reduction criteria, each meeting a different tradeoff between quality of the reduced basis and the computational effort required for finding it. The most popular one is the Lenstra-Lenstra-Lovasz (LLL) reduction [30], which can be interpreted as an extension of the Gauss reduction to lattices of rank greater than 2.

The rest of this paper is organized as follows. Section II introduces the necessary notation and formulates the transform identification problem and Section III provides the background on lattice theory. The proposed method is described in Section IV. Then, a theoretical analysis of the convergence properties is presented in Section V. The performance of the transform identification algorithm is evaluated empirically in Section VI. Finally, Section VII concludes the paper, indicating the open issues and stimulating further investigations.

## II. PROBLEM STATEMENT

The symbols $x$, $\mathbf{x}$ and $\mathbf{X}$ denote, respectively, a scalar, a column vector and a matrix. A $M \times N$ matrix $\mathbf{X}$ can be written either in terms of its columns or rows. Specifically,

$$\mathbf{X} = \begin{bmatrix} \mathbf{x}_1 & \mathbf{x}_2 & \ldots & \mathbf{x}_N \end{bmatrix} = \begin{bmatrix} \bar{\mathbf{x}}_1^T \\ \bar{\mathbf{x}}_2^T \\ \ldots \\ \bar{\mathbf{x}}_M^T \end{bmatrix}. \quad (1)$$

Let $\mathbf{x}$ denote a $N$-dimensional vector and $\mathbf{W}$ a transform matrix, whose rows represent the transform basis functions.

Transform coding is performed by applying scalar quantization to the transform coefficients $\mathbf{y} = \mathbf{W}\mathbf{x}$. Let $\mathcal{Q}_i(\cdot)$ denote the quantizer associated to the $i$-th transform coefficient. We assume that $\mathcal{Q}_i(\cdot)$ is a scalar uniform quantizer with step size $\Delta_i$, $i = 1, \ldots, N$. Therefore, the reconstructed quantized coefficients can be written as $\tilde{\mathbf{y}} = [\tilde{y}_1, \tilde{y}_2, \ldots, \tilde{y}_N]^T$, with

$$\tilde{y}_i = \mathcal{Q}_i(y_i) = \Delta_i \cdot \text{round}\left[\frac{y_i}{\Delta_i}\right], \quad i = 1, \ldots, N. \tag{2}$$

The reconstructed block in the original domain is given by $\tilde{\mathbf{x}} = \mathbf{W}^{-1}\tilde{\mathbf{y}}$.

Let $\{\tilde{\mathbf{x}}_1, \ldots, \tilde{\mathbf{x}}_P\}$ denote a set of $P$ observed $N$-dimensional vectors, which are the output of a transform coder. Due to quantization, the unobserved vectors representing quantized transform coefficients $\{\tilde{\mathbf{y}}_1, \ldots, \tilde{\mathbf{y}}_P\}$ are constrained to belong to a lattice $\mathcal{L}_y$ described by the following basis:

$$\mathbf{B}_y = \begin{bmatrix} \Delta_1 & 0 & \ldots & 0 \\ 0 & \Delta_2 & \ldots & 0 \\ \vdots & \vdots & \ddots & \vdots \\ 0 & 0 & \ldots & \Delta_N \end{bmatrix} \tag{3}$$

Therefore, the observed vectors $\{\tilde{\mathbf{x}}_1, \ldots, \tilde{\mathbf{x}}_P\}$ belong to a lattice $\mathcal{L}_x$ described by the basis:

$$\mathbf{B}_x = [\mathbf{b}_{x,1}, \ldots, \mathbf{b}_{x,N}] = \mathbf{W}^{-1}\mathbf{B}_y, \tag{4}$$

with $\mathbf{b}_{x,i} = \Delta_i \hat{\mathbf{w}}_i$, $i = 1, \ldots, N$, $\mathbf{W}^{-1} = [\hat{\mathbf{w}}_1, \ldots, \hat{\mathbf{w}}_N]$.

In this paper we study the problem of determining $\mathbf{B}_x$ from a finite set of $P \geq N$ distinct vectors $\{\tilde{\mathbf{x}}_1, \ldots, \tilde{\mathbf{x}}_P\}$. That is, we seek to determine the parameters of a transform coder based on the footprints left on its output. We propose an algorithm to solve this problem and we study its convergence properties. In addition, we show that the probability of correctly determining $\mathbf{B}_x$ (or, equivalently, another basis for the lattice $\mathcal{L}_x$) is monotonically increasing in the number of observations $P$, and rapidly approaching one when $P > N$. Note that when determining $\mathbf{B}_x$, the proposed method does not make any assumption on the structure of the transform matrix $\mathbf{W}$. In the general case, given $\mathbf{B}_x$, it is not possible to uniquely determine $\mathbf{W}$ and the quantization step sizes $\Delta_i$, $i = 1, \ldots, N$. Indeed, the length of each basis vector $\mathbf{b}_{x,i}$ can be factored out as $\|\mathbf{b}_{x,i}\|_2 = \Delta_i \|\hat{\mathbf{w}}_i\|_2$. However, in the important case in which $\mathbf{W}$ represents an orthonormal transform, the quantization step sizes $\Delta_i$, $i = 1, \ldots, N$, and the transform matrix $\mathbf{W}$ can be immediately obtained from $\mathbf{B}_x$. Indeed, $\mathbf{W}^{-1} = \mathbf{W}^T$, $\hat{\mathbf{w}}_i = \bar{\mathbf{w}}_i$, $i = 1, \ldots, N$, with $\|\bar{\mathbf{w}}_i\|_2 = 1$. Therefore:

$$\Delta_i = \|\mathbf{b}_{x,i}\|_2, \quad i = 1, \ldots, N, \tag{5}$$

$$\bar{\mathbf{w}}_i = \mathbf{b}_{x,i}/\|\mathbf{b}_{x,i}\|_2 \quad i = 1, \ldots, N. \tag{6}$$

## III. BACKGROUND ON LATTICE THEORY

In this section we provide the necessary background on lattice theory. Further details can be found, e.g., in [31][32][28]. Let $\mathcal{L}$ denote a lattice of rank $K$ embedded in $\mathbb{R}^N$. Let $\mathbf{B} = [\mathbf{b}_1, \mathbf{b}_2, \ldots, \mathbf{b}_K]$ denote a basis for the lattice $\mathcal{L}$. That is,

$$\mathcal{L} = \{\mathbf{x} | a_1 \mathbf{b}_1 + a_2 \mathbf{b}_2 + \ldots + a_K \mathbf{b}_K, a_i \in \mathbb{Z}\}. \tag{7}$$

In order to make the mapping between a basis and the corresponding lattice explicit, the latter can be expressed as $\mathcal{L}(\mathbf{B})$.

Any lattice basis also describes a fundamental parallelotope according to

$$\mathcal{P}(\mathbf{B}) = \left\{ \mathbf{x} | \mathbf{x} = \sum_{i=1}^{K} \theta_i \mathbf{b}_i, 0 \leq \theta_i < 1 \right\}. \tag{8}$$

When $K = 2, 3$, $\mathcal{P}(\mathbf{B})$ is, respectively, a parallelogram or a parallelepiped. As an example, Figure 1(a) shows the fundamental parallelotope corresponding to a lattice basis $\mathbf{B}$ when $K = 2$.

Given a point $\mathbf{z} \in \mathbb{R}^K$, let $\mathcal{P}_{\mathbf{z}}(\mathbf{B})$ denote the parallelotope enclosing $\mathbf{z}$. $\mathcal{P}_{\mathbf{z}}(\mathbf{B})$ is obtained by translating $\mathcal{P}(\mathbf{B})$ so that its origin coincides with one of the lattice points. More specifically,

$$\mathcal{P}_z(\mathbf{B}) = \left\{ \mathbf{x} | \mathbf{x} = \mathbf{B} \cdot \lfloor \mathbf{B}^{-1} \mathbf{z} \rfloor + \sum_{i=1}^{K} \theta_i \mathbf{b}_i, 0 \leq \theta_i < 1 \right\}. \tag{9}$$

Figure 1(b) illustrates $\mathcal{P}_{\mathbf{z}}(\mathbf{B})$ for an arbitrary vector $\mathbf{z}$.

Different bases for the same lattice lead to different fundamental parallelotopes. For example, Figure 1(a) and Figure 1(c) depict two different bases for the same lattice, together with the corresponding fundamental parallelotopes. However, the volume of $\mathcal{P}(\mathbf{B})$ is the same for all bases of a given lattice. This volume equals the so-called *lattice determinant*, which is a lattice invariant defined as

$$|\mathcal{L}| = \sqrt{\det(\mathbf{B}^T \mathbf{B})}. \tag{10}$$

If the lattice is full rank, i.e., $K = N$, the lattice determinant equals the determinant of the matrix $\mathbf{B}$, $|\mathcal{L}| = |\det(\mathbf{B})|$.

Let $\underline{\mathcal{L}}$ denote a sub-lattice of $\mathcal{L}$. That is, for any vector $\mathbf{x} \in \underline{\mathcal{L}}$, then $\mathbf{x} \in \mathcal{L}$. A basis $\underline{\mathbf{B}}$ for $\underline{\mathcal{L}}$ can be expressed in terms of $\mathbf{B}$ as

$$\underline{\mathbf{B}} = \mathbf{B}\mathbf{U}, \tag{11}$$

where $\mathbf{U}$ is such that $u_{ij} \in \mathbb{Z}$. Moreover, let $\det(\mathbf{U}) = \pm m$, then

$$\frac{|\underline{\mathcal{L}}|}{|\mathcal{L}|} = |\det(\mathbf{U})| = m \tag{12}$$

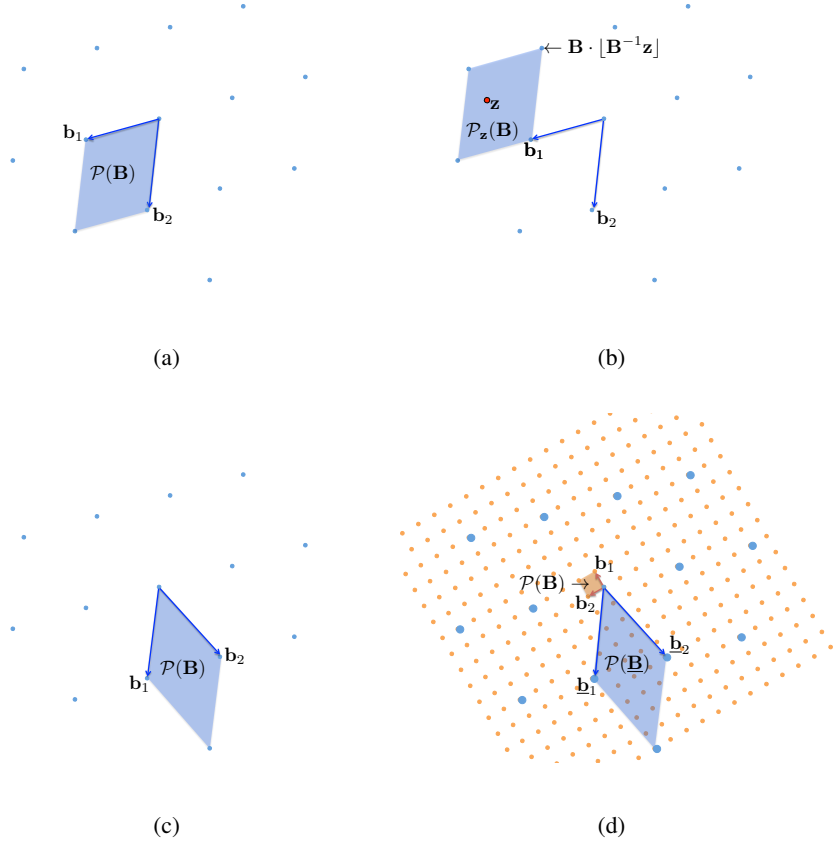

Fig. 1. Examples of lattices. (a) The fundamental parallelotope of a lattice defined by a basis $\mathbf{B}$. (b) Parallelotope enclosing an arbitrary vector $\mathbf{z}$. (c) Another (equivalent) basis for the lattice in (a). (d) An example of a sub-lattice of the lattice $\mathcal{L}(\mathbf{B})$.

and we say that $\underline{\mathcal{L}}$ is a sub-lattice of $\mathcal{L}$ of index $m$. For example, Figure 1(d) shows two lattices $\underline{\mathcal{L}}$ and $\mathcal{L}$, such that $\underline{\mathcal{L}} \subset \mathcal{L}$. In this case, the matrix $\mathbf{U}$ is equal to

$$\mathbf{U} = \begin{bmatrix} -4 & -5 \\ 3 & -1 \end{bmatrix}, \quad (13)$$

and $\underline{\mathcal{L}}$ is a sub-lattice of index $m = 19$.

## IV. AN ALGORITHM FOR TRANSFORM IDENTIFICATION

In this section we propose an algorithm that is able to determine the parameters of a transform coder from its output, i.e., a set of observed vectors $\{\tilde{\mathbf{x}}_1, \ldots, \tilde{\mathbf{x}}_P\}$. This is accomplished by finding a suitable lattice $\mathcal{L}^*$ such that $\{\tilde{\mathbf{x}}_1, \ldots, \tilde{\mathbf{x}}_P\} \subset \mathcal{L}^*$. In Section V-C we will show that, with probability approaching one, $\mathcal{L}^* \equiv \mathcal{L}_x$, provided that $P - N > 0$.

The problem of determining a basis for the lattice $\mathcal{L}_x$ is complicated by the fact that we typically observe a finite (and possibly small) number of vectors $P$ embedded in a possibly large dimensional space. More precisely, $\{\tilde{\mathbf{x}}_1, \ldots, \tilde{\mathbf{x}}_P\}$ belong to a bounded lattice, in virtue of the fact that each transform coefficient $y_i$ is quantized with a finite number of bits $R_i$, to one of $2^{R_i}$ reconstruction levels. Let $\bar{R}$ denote the average number of bits allocated to transform coefficients. The number of potential lattice points is equal to

$$\prod_{i=1}^{N} 2^{R_i} = 2^{\sum_{i=1}^{N} R_i} = 2^{N\bar{R}}, \tag{14}$$

and only $P$ of them are covered by observed vectors. Thus, we note that, given $\bar{R}$, the number of lattice points increases exponentially with the dimension $N$ and that in most cases of practical relevance $P \ll 2^{N\bar{R}}$.

Another issue arises from the fact that, for a set of vectors $\{\tilde{\mathbf{x}}_1, \ldots, \tilde{\mathbf{x}}_P\}$, there are infinitely many lattices that include all of them. Indeed, any lattice $\bar{\mathcal{L}}$ such that $\mathcal{L}_x \subset \bar{\mathcal{L}}$ is compatible with the observed set of vectors. Note that any basis of the form $\mathbf{B} = \mathbf{B}_x \mathbf{U}^{-1}$, with $\det(\mathbf{U}) = \pm m$, with $m$ an integer greater than one defines a compatible lattice $\bar{\mathcal{L}}$. A simple example is obtained setting $\mathbf{U} = a\mathbf{I}$, $a \in \mathbb{N}, a > 1$.

In order to resolve this ambiguity, we seek the lattice $\mathcal{L}^*$ that maximizes the lattice determinant $|\mathcal{L}|$, within this infinite set of compatible lattices. That is,

$$\begin{aligned} \underset{\mathcal{L}(\mathbf{B})}{\text{maximize}} \quad & |\mathcal{L}(\mathbf{B})| \\ \text{subject to} \quad & \{\tilde{\mathbf{x}}_1, \ldots, \tilde{\mathbf{x}}_P\} \subset \mathcal{L}(\mathbf{B}). \end{aligned} \tag{15}$$

For example, for the set of observed points $\{\tilde{\mathbf{x}}_1, \tilde{\mathbf{x}}_2, \tilde{\mathbf{x}}_3\}$ depicted in Figure 2(a), Figure 2(g) illustrates a basis for the lattice that is the optimal solution of (15). In contrast, the lattice in Figure 2(h) is a feasible solution of (15), but it is not optimal, since it is characterized by a lower value of the lattice determinant.

The proposed method used to solve the problem above is detailed in Algorithm 1. The method constructs an initial basis for an $N$-dimensional lattice (line 1). This is accomplished by considering the vectors in $\mathcal{O}$ until $N$ linearly independent vectors are found. These vectors are used as columns of the starting estimate $\mathbf{B}^{(0)}$ and to populate the initial set of visited vectors $\mathcal{S}$. We denote with $\mathcal{U}$ the set of vectors in $\mathcal{O}$ that have not been visited yet. Then, the solution of (15) is constructed iteratively, by considering the remaining vectors in $\mathcal{U}$ one by one. At each iteration, the function `recurseTI` returns a basis for a lattice that solves (15), in which the constraint is imposed only on the subset of visited vectors $\mathcal{S}$, that

**ALGORITHM 1:** `TI` algorithm

Input: *Set of observed vectors* $\mathcal{O} = \{\tilde{\mathbf{x}}_1, \ldots, \tilde{\mathbf{x}}_P\}$

Output: *A basis* $\mathbf{B}$ *of the lattice solution of* (15)

1) $\mathbf{B}^{(0)} = \text{initBasis}(\mathcal{O})$;
2) $\mathcal{S} = \{\mathbf{b}_1, \ldots, \mathbf{b}_N\}$;
3) $\mathcal{U} = \mathcal{O} \setminus \mathcal{S}$;
4) $r = 0$
5) **while** $\text{card}\{\mathcal{U}\} > 0$;
6)    Pick $\tilde{\mathbf{x}}$ in $\mathcal{U}$;
7)    $\mathcal{U} = \mathcal{U} \setminus \{\tilde{\mathbf{x}}\}$;
8)    $\mathcal{S} = \mathcal{S} \cup \tilde{\mathbf{x}}$;
9)    $\mathbf{B}^{(r+1)} = \text{recurseTI}(\mathbf{B}^{(r)}, \mathcal{S})$;
10)   $r = r + 1$
11) **end**

is, $\mathcal{S} \subset \mathcal{L}(\mathbf{B})$. As such, the algorithm starts finding the solution of an under-constrained problem and additional constraints are added as more vectors are visited.

Figure 2 shows an illustrative example when $N = 2$ and three vectors $\{\tilde{\mathbf{x}}_1, \tilde{\mathbf{x}}_2, \tilde{\mathbf{x}}_3\}$ are observed (Figure 2(b)). The initial basis (line 1) is constructed using $\tilde{\mathbf{x}}_1$ and $\tilde{\mathbf{x}}_2$, since they are linearly independent (Figure 2(b)). Then, the point $\tilde{\mathbf{x}}_3$ is selected (line 6 and Figure 2(c)) and the function `recurseTI` (line 9) returns a basis that solves (15), i.e., a basis with the largest lattice determinant that includes all observed vectors. Figure 2(f) illustrates such a basis, and Figure 2(g) shows an equivalent basis obtained after lattice reduction.

The core of the method is the recursive function `recurseTI`. When describing this function, we keep a clear distinction between algorithm template and algorithm instance, as it is customary in computer science. We start describing the template in Algorithm 2, which does not specify the function entirely. Then, a concrete instance of the template is detailed in Algorithm 3. The rationale of maintaining this distinction is motivated by the fact that the correctness of the method is a property that descends from the template alone, as further discussed in Section V-A. Conversely, the rate of convergence depends on the specific algorithm instance, as explained in Section V-B.

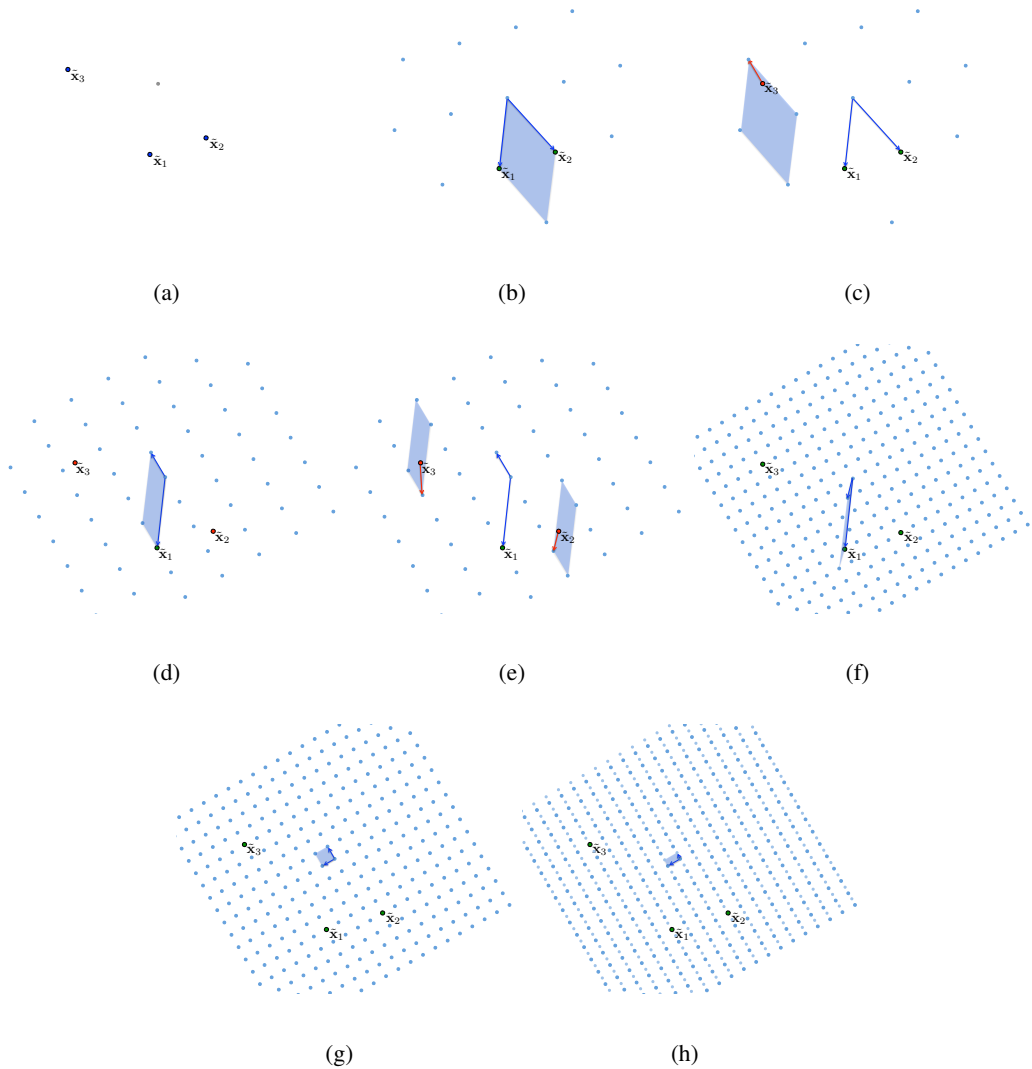

Fig. 2. An example of transform identification. A set of three observed vectors is given in (a). Then, (b)-(h) show, step-by-step, how the solution to problem (15) is sought by Algorithm 1.

## A. An algorithm template for `recurseTI`

The function `recurseTI` receives as input a set of visited vectors $\mathcal{S}$ and the current estimate of a basis $\mathbf{B}$ for the lattice $\mathcal{L}(\mathbf{B})$. If $\mathcal{S} \subset \mathcal{L}$, i.e., all the vectors in $\mathcal{S}$ belong to the lattice defined by $\mathbf{B}$, the recursion is terminated (line 1 in Algorithm 2). Otherwise, one of the vectors $\tilde{\mathbf{x}}$ that does not belong to $\mathcal{L}$ is selected (line 4) and the parallelotope which encloses it is identified (line 5). Then, a vector $\mathbf{d}$ is computed as the difference between $\tilde{\mathbf{x}}$ and one of the vertices of the parallelotope (line 6). The intuition

**ALGORITHM 2:** `recurseTI(B, S)`

Input: *Set of vectors $S = \{\tilde{x}_1, \ldots, \tilde{x}_S\}$, a basis $B$ of a lattice.*

Output: *A basis of a lattice $\mathcal{L}$ with maximum determinant $|\mathcal{L}|$, such that $S \subset \mathcal{L}$*

1)     **if** $S \subset \mathcal{L}(B)$
2)         **return $B$**
3)     **else**
4)         Pick $z \in S \setminus \mathcal{L}(B)$.
5)         Determine $\mathcal{P}_z(B)$.
6)         Pick a vertex $v$ of $\mathcal{P}_z(B)$.
7)         Compute $d = z - v$.
8)         Compute $B_i$, replacing the $i$-th column of $B$ with $d$.
9)         Pick an index $l$, such that $\det(B_l) \neq 0$.
10)        `recurseTI`$(B_l, S)$;
11)     **end**

here is to capture a short vector that cannot be represented by the current lattice, and to modify the current basis in such a way that (upon convergence) it can be represented. Hence, the updated basis is constructed by replacing one of the columns of $B$ with $d$ (line 8). Among the $N$ possible cases, any choice such that $B_i$ is non-singular represents a valid selection (line 10).

In the example in Figure 2, two recursive steps are performed before terminating `recurseTI`. In the first call, it is verified that $\tilde{x}_3$ does not belong to the lattice defined by the current basis (Figure 2(c)), and the updated basis is constructed (Figure 2(d)) by replacing one of the two basis vectors with the difference vector between $\tilde{x}_3$ and one of the vertices of $\mathcal{P}_{\tilde{x}_3}(B)$. In the second call it is verified that neither $\tilde{x}_3$ nor $\tilde{x}_2$ belong to the updated lattice. Therefore, one of the two difference vectors (e.g., the one representing the difference between $\tilde{x}_2$ and one of the vertices of $\mathcal{P}_{\tilde{x}_2}(B)$) is used to replace one of the two basis vectors. In the third call the recursion is terminated, because all points in $S$ belong to the lattice.

In Section V-A, it is shown that the recursion always terminates in a finite number of steps and leads to the optimal solution of (15). The solution the algorithm converges to, though, might be a sub-lattice of the underlying lattice $\mathcal{L}_x$, i.e., $\mathcal{L}^* \subset \mathcal{L}_x$. Fortunately, this is a very unlikely event, even when the number of observed points $P$ is only slightly larger than $N$, as further discussed in Section V-C.

## B. An algorithm instance for `recurseTI`

A practical instantiation of the template presented in Algorithm 2 requires to specify how to perform the choices at line 4, 6 and 9, which were left undefined. Note that these choices are arbitrary and have no effect on the correctness of the method, although they might affect the number of recursive steps needed to achieve convergence.

In our specific implementation, the selection of the vector $\tilde{\mathbf{x}} \in \mathcal{S} \setminus \mathcal{L}(\mathbf{B})$ (line 4 in Algorithm 2), the vertex of the parallelotope (line 6) and the column to be replaced (line 9) are carried out as detailed in Algorithm 3. The rationale is to construct a new basis related to a lattice with the smallest lattice determinant $|\mathcal{L}(\mathbf{B})|$, so as to tighten the upper bound on the value of the optimal solution, i.e., $|\mathcal{L}^*| \leq |\mathcal{L}(\mathbf{B})|$.

Specifically, given a basis $\mathbf{B}$ as input, we compute the vector $\hat{\mathbf{x}} = \mathbf{B} \cdot \text{round}(\mathbf{B}^{-1}\tilde{\mathbf{x}})$, which represents one of the vertices of the parallelotope enclosing $\tilde{\mathbf{x}}$ (line 4 in Algorithm 3). In order to prevent numerical instability induced by the inversion of the matrix $\mathbf{B}$, we perform basis reduction according to the LLL algorithm (line 2) and we find a nearly orthogonal basis which is equivalent to $\mathbf{B}$, but has a smaller orthogonality defect. In our implementation, we perform basis reduction only when the condition number is greater than a threshold $T$, which was set equal to $10^4$ (line 1).

Then, the selected point $\mathbf{z} = \tilde{\mathbf{x}}_f$ is the one that minimizes the distance from the corresponding vertex (line 8). That is,

$$f = \arg \min_{j \in \{l | \|\tilde{\mathbf{x}}_l - \hat{\mathbf{x}}_l\|_2 > 0\}} \|\tilde{\mathbf{x}}_j - \hat{\mathbf{x}}_j\|_2, \tag{16}$$

so as to minimize the length of the new basis vector $\mathbf{d}$. Similarly, the choice of the new basis among the set of (up to) $N$ candidate bases $\mathbf{B}_i$ (line 11) is to select the one that leads to the smallest lattice determinant, after excluding those that do not have rank $N$. From Cramer's rule, it follows that $\det(\mathbf{B}_i) = \theta_i \det(\mathbf{B})$, where $\boldsymbol{\theta} = \mathbf{B}^{-1}\mathbf{d}$ is the expansion of $\mathbf{d}$ in the basis $\mathbf{B}$. Hence, we replace the $l$-th column of $\mathbf{B}$, which is the one corresponding to the entry of $\boldsymbol{\theta}$ with the least strictly positive absolute value. That is,

$$l = \arg \min_{j \in \{p | \theta_p \neq 0\}} |\theta_j|. \tag{17}$$

## V. ANALYSIS

### A. Convergence

In this section, we prove that the proposed algorithm converges in a finite number of recursive steps to the solution $\mathcal{L}^*$ of (15). To this end, we rely on the specifications of the algorithm template in Algorithm 2.

**ALGORITHM 3:** `recurseTI(B, S)`

Input: *Set of vectors $\mathcal{S} = \{\tilde{\mathbf{x}}_1, \ldots, \tilde{\mathbf{x}}_S\}$, a basis $\mathbf{B}$ of a lattice.*

Output: *A basis of a lattice $\mathcal{L}$ with maximum determinant $|\mathcal{L}|$, such that $\mathcal{S} \subset \mathcal{L}$*

1) **if** condnum($\mathbf{B}$) > $T$
2)     $\mathbf{B} = \text{LLL}(\mathbf{B})$
3) **end**
4) $\hat{\mathbf{x}}_i = \mathbf{B} \cdot \text{round}(\mathbf{B}^{-1}\tilde{\mathbf{x}}_i)$, $i = 1, \ldots, S$;
5) **if** $(\max_{j=1,\ldots,S} \|\tilde{\mathbf{x}}_j - \hat{\mathbf{x}}_j\|_2) = 0$
6)     **return** $\mathbf{B}$
7) **else**
8)     $f = \arg\min_{j \in \{l | \|\tilde{\mathbf{x}}_l - \hat{\mathbf{x}}_l\|_2 > 0\}} \|\tilde{\mathbf{x}}_j - \hat{\mathbf{x}}_j\|_2$;
9)     $\mathbf{d} = \tilde{\mathbf{x}}_f - \hat{\mathbf{x}}_f$;
10)     $\boldsymbol{\theta} = \mathbf{B}^{-1}\mathbf{d}$;
11)     $l = \arg\min_{j \in \{p | \theta_p \neq 0\}} |\theta_j|$;
12)     `recurseTI`($\mathbf{B}_l, \mathcal{S}$);
13) **end**

Let $\mathbf{B}^{(0)}$ denote the initial estimate of a basis of the lattice, which is constructed, for example, by selecting as its columns a subset of $N$ linearly independent vectors in $\mathcal{O}$ (Algorithm 1, line 1). Hence, each vector of the initial basis $\mathbf{B}^{(0)}$ can be expressed as a linear combination with integer coefficients of the columns of $\mathbf{B}_x$. Thus, we can write $\mathbf{B}^{(0)} = \mathbf{B}_x \mathbf{A}$, with $\det(\mathbf{A}) = m$ and $m \in \mathbb{Z} \setminus \{0\}$. From this, it follows that $|\mathcal{L}(\mathbf{B}^{(0)})| = m \cdot |\mathcal{L}_x|$ and $|\mathcal{L}_x| \leq |\mathcal{L}(\mathbf{B}^{(0)})|$

Let $\mathbf{B}^{(r)}$ denote the estimate obtained after the $r$-th call of the recursive function `recurseTI`. It is possible to prove the following lemma:

*Lemma 5.1:* $|\mathcal{L}(\mathbf{B}^{(r+1)})| \leq |\mathcal{L}(\mathbf{B}^{(r)})|$, with equality if and only if $\mathcal{S} \subset \mathcal{L}(\mathbf{B}^{(r)}) = \mathcal{L}(\mathbf{B}^{(r+1)})$

*Proof:* If $\mathcal{S} \subset \mathcal{L}(\mathbf{B}^{(r)})$, then $\mathbf{B}^{(r+1)} = \mathbf{B}^{(r)}$ and the recursion terminates. Otherwise, let $\mathbf{z} \in \mathcal{S} \setminus \mathcal{L}(\mathbf{B}^{(r)})$ be any of the points which does not belong to the lattice defined by $\mathbf{B}^{(r)}$, $\mathbf{v}$ any of the vertices of $\mathcal{P}_{\mathbf{z}}(\mathbf{B}^{(r)})$ and $\mathbf{d} = \mathbf{z} - \mathbf{v}$. The vector $\mathbf{d}$ can be expressed in terms of the basis $\mathbf{B}^{(r)}$ as

$$\mathbf{d} = \mathbf{B}^{(r)}\boldsymbol{\theta}. \tag{18}$$

By definition, the vector $\mathbf{z}$ belongs to $\mathcal{P}_{\mathbf{z}}(\mathbf{B}^{(r)})$, hence $-1 \leq \theta_i \leq 1$. Since $\mathbf{z} \notin \mathcal{L}(\mathbf{B}^{(r)})$, $\mathbf{z}$ does not belong to the vertices of $\mathcal{P}_{\mathbf{z}}(\mathbf{B}^{(r)})$. It follows that there is at least one coefficient $\theta_l$ in the basis expansion of $\mathbf{d}$, such that $0 < |\theta_l| < 1$.

The vector $\mathbf{d}$ replaces the $i$-th column of $\mathbf{B}^{(r)}$ to obtain $\mathbf{B}_i^{(r)}$. From Cramer's rule,

$$\det(\mathbf{B}_i^{(r)}) = \theta_i \det(\mathbf{B}^{(r)}) \tag{19}$$

Therefore, if we select $l$, such that $0 < |\theta_l| < 1$,

$$|\mathcal{L}(\mathbf{B}^{(r+1)})| = |\det(\mathbf{B}^{(r+1)})| = |\det(\mathbf{B}_l^{(r)})| = |\theta_l||\det(\mathbf{B}^{(r)})| < |\det(\mathbf{B}^{(r)})| = |\mathcal{L}(\mathbf{B}^{(r)})| \tag{20}$$

Note that there must be at least one such an index $l$, as indicated above.

∎

We construct the sequence of integer numbers

$$s_r = |\mathcal{L}(\mathbf{B}^{(r)})|, \quad r = 0, 1, \ldots, R. \tag{21}$$

Let $R$ denote the smallest integer such that $|\mathcal{L}(\mathbf{B}^{(R)})| = |\mathcal{L}(\mathbf{B}^{(R+1)})|$. That is, $R$ is the number of steps needed to achieve convergence. It is possible to prove the following theorem:

*Theorem 5.2:* Algorithm 1 converges to the solution of (15).

*Proof:* Let $\mathcal{L}^*$ denote the solution of (15), i.e., the lattice with maximum volume that includes all observed vectors $\mathcal{S}$. We need to prove that $\mathcal{L}(\mathbf{B}^{(R)}) = \mathcal{L}^*$.

First, we prove that $|\mathcal{L}(\mathbf{B}^{(R)})|$ cannot decrease beyond $|\mathcal{L}^*|$, i.e., $|\mathcal{L}^*| \leq |\mathcal{L}(\mathbf{B}^{(R)})|$. To this end, let $\mathcal{L}(\mathbf{B}^{(R-1)})$ denote the lattice obtained at the iteration just before convergence. Hence, there is at least one observed vector $\tilde{\mathbf{x}} \in \mathcal{L}^*$ such that $\tilde{\mathbf{x}} \notin \mathcal{L}(\mathbf{B}^{(R-1)})$. Lemma 5.1 establishes that $|\mathcal{L}(\mathbf{B}^{(R)})| < |\mathcal{L}(\mathbf{B}^{(R-1)})|$.

Let $\mathbf{d}$ denote the difference vector as in line 7 of Algorithm 2. By construction, $\mathbf{d} \in \mathcal{L}^*$. Let $\mathbf{B}^*$ denote a basis for $\mathcal{L}^*$. Then, it is possible to write $\mathbf{d} = \mathbf{B}^* \boldsymbol{\theta}^*$, $\theta_i^* \in \mathbb{Z}$. $\mathcal{L}(\mathbf{B}^{(R-1)})$ is a sublattice of $\mathcal{L}^*$. Hence, $\mathbf{B}^{(R-1)} = \mathbf{B}^* \mathbf{A}$, where $\mathbf{A}$ is a matrix of integer elements such that $\det(\mathbf{A}) = m$, with $m \in \mathbb{Z} \setminus \{0\}$, and $|\mathcal{L}(\mathbf{B}^{(R-1)})|/|\mathcal{L}^*| = m$.

It is possible to express $\mathbf{d}$ in the basis expansion of $\mathbf{B}^{(R-1)}$. That is,

$$\boldsymbol{\theta} = (\mathbf{B}^{(R-1)})^{-1}\mathbf{d} = (\mathbf{B}^*\mathbf{A})^{-1}\mathbf{B}^*\boldsymbol{\theta}^* = \mathbf{A}^{-1}\boldsymbol{\theta}^* = \frac{1}{\det(\mathbf{A})}\text{cofactor}(\mathbf{A})\boldsymbol{\theta}^*. \tag{22}$$

Note that both the cofactor matrix $\text{cofactor}(\mathbf{A})$ and $\boldsymbol{\theta}^*$ have integer elements. Hence, the vector $\text{cofactor}(\mathbf{A})\boldsymbol{\theta}^*$ has integer elements. Any nonzero element of $\boldsymbol{\theta}$ is an integer multiple of $1/\det(\mathbf{A}) = 1/m$. Therefore, if $\theta_i \neq 0$, $|\theta_i| \geq 1/m$.

From the proof of Lemma 5.1, we know that

$$|\mathcal{L}(\mathbf{B}^{(R)})| = |\theta_l||\mathcal{L}(\mathbf{B}^{(R-1)})| \geq \frac{1}{m}|\mathcal{L}(\mathbf{B}^{(R-1)})| = |\mathcal{L}^*|, \tag{23}$$

where $\theta_l$ is one of the nonzero elements of $\boldsymbol{\theta}$.

To prove that $|\mathcal{L}(\mathbf{B}^{(R)})| = |\mathcal{L}^*|$, it remains to be shown that cannot be $|\mathcal{L}(\mathbf{B}^{(R)})| > |\mathcal{L}^*|$. Indeed, if this were the case, $\mathcal{L}(\mathbf{B}^{(R)})$ would be the optimal solution of (15), since it includes all observed points $\mathcal{S}$ and has volume larger than $|\mathcal{L}^*|$. ∎

Note that $R < \infty$, i.e., convergence is achieved in a finite number of steps. Indeed, $\{s_r\}$ is a sequence of integer values. The sequence is monotonically decreasing due to Lemma 5.1, until convergence is achieved and $\mathcal{S} \subset \mathcal{L}(\mathbf{B}^{(R)})$. In addition, it is bounded from below by $|\mathcal{L}_x|$. Therefore, convergence is achieved in up to $|\mathcal{L}(\mathbf{B}^{(0)})|/|\mathcal{L}_x|$ number of steps. In the following section we show that with a specific instantiation of Algorithm 2 given in Algorithm 3 it is possible to ensure a significantly faster convergence rate.

## B. Rate of convergence

It is possible to prove that the proposed method implemented according to the instance presented in Algorithm 3 converges in a number of steps that is upper bounded by $\lceil \log_2(|\mathcal{L}(\mathbf{B}^{(0)})|/|\mathcal{L}_x|) \rceil$. To show this, it suffices to demonstrate that the value of the lattice determinant is (at least) halved between two consecutive calls of `recurseTI`, as stated by the following theorem.

*Theorem 5.3:* If $\mathcal{S} \not\subset \mathcal{L}(\mathbf{B}^{(r)})$, then $\frac{|\mathcal{L}(\mathbf{B}^{(r+1)})|}{|\mathcal{L}(\mathbf{B}^{(r)})|} \leq \frac{1}{2}$

*Proof:* Since $\mathcal{S} \not\subset \mathcal{L}(\mathbf{B}^{(r)})$, then $\max_{j=1,\ldots,S} \|\tilde{\mathbf{x}}_j - \hat{\mathbf{x}}_j\|_2 > 0$, and the recursion is not terminated. Consider the vector $\mathbf{d} = \tilde{\mathbf{x}}_f - \hat{\mathbf{x}}_f$, which can be expressed in the basis $\mathbf{B}^{(r)}$ as $\mathbf{d} = \mathbf{B}^{(r)}\boldsymbol{\theta}$. Dropping the superscript $^{(r)}$, it is possible to write

$$\boldsymbol{\theta} = \mathbf{B}^{-1}\mathbf{d} = \mathbf{B}^{-1}(\tilde{\mathbf{x}}_f - \hat{\mathbf{x}}_f) \tag{24}$$

$$= \mathbf{B}^{-1}\tilde{\mathbf{x}}_f - \mathbf{B}^{-1}(\mathbf{B} \cdot \text{round}(\mathbf{B}^{-1}\hat{\mathbf{x}}_f)) \tag{25}$$

$$= \mathbf{B}^{-1}\tilde{\mathbf{x}}_f - \text{round}(\mathbf{B}^{-1}\hat{\mathbf{x}}_f) = \mathbf{a} - \text{round}(\mathbf{a}), \tag{26}$$

where we set $\mathbf{a} = \mathbf{B}^{-1}\hat{\mathbf{x}}_f$. Due to the properties of rounding, $-1/2 \leq \theta_i < 1/2$. Thus, replacing any of the columns of $\mathbf{B}^{(r)}$ such that $\theta_l \neq 0$, we obtain, using Cramer's rule,

$$\frac{|\mathcal{L}(\mathbf{B}^{(r+1)})|}{|\mathcal{L}(\mathbf{B}^{(r)})|} = |\theta_l| < \frac{1}{2} \tag{27}$$

∎

Based on Theorem 5.3,

$$|\mathcal{L}(\mathbf{B}^{(r)})| \leq \left(\frac{1}{2}\right)^r |\mathcal{L}(\mathbf{B}^{(0)})|, \quad \forall r > 0, \mathcal{S} \not\subset \mathcal{L}(\mathbf{B}^{(r)}) \tag{28}$$

Hence, convergence is achieved in up to

$$\left\lceil \log_2 \frac{|\mathcal{L}(\mathbf{B}^{(0)})|}{|\mathcal{L}_x|} \right\rceil \tag{29}$$

number of steps.

Note that this upper bound on the convergence rate is guaranteed solely on the basis of the way the vertex of the parallelotope is selected, whereas it does not depend neither on which point is selected, nor on which column is replaced. However, the heuristics applied in Algorithm 3 are based on the rationale of reducing the ratio $\frac{|\mathcal{L}(\mathbf{B}^{(r+1)})|}{|\mathcal{L}(\mathbf{B}^{(r)})|}$ as much as possible.

## C. Probability of success

In Section V-A, we showed that the proposed method converges to the optimal solution $\mathcal{L}^*$ of (15). In this section, we show that it converges to the correct (and unique) lattice $\mathcal{L}_x$ (i.e., $\mathcal{L}^* \equiv \mathcal{L}_x$) with high probability, provided that the number of observed vectors $P$ is greater than $N$.

Given a lattice $\mathcal{L}_x$ of rank $N$ embedded in $\mathbb{R}^N$, there is more than one sub-lattice $\underline{\mathcal{L}}$ of $\mathcal{L}$ of index $m$. It can be shown that the number of sub-lattices is equal to [33]

$$f_N(m) = \prod_{i=1}^{q} \prod_{j=1}^{N-1} \frac{p_i^{t_i+j} - 1}{p_i^j - 1} = \prod_{i=1}^{q} \prod_{j=1}^{t_i} \frac{p_i^{N+j-1} - 1}{p_i^j - 1}, \tag{30}$$

where $m = p_1^{t_1} \cdots p_q^{t_q}$ is the prime factorization of $m$. That is, $p_1, \ldots, p_q$ are the prime factors of $m$, and $t_s$ is the multiplicity of the factor $p_s$.

For example, when $N = 2$ and $m = 2$, $f_2(2) = 3$. Given the basis $\mathbf{B} = \mathbf{I}$, the corresponding sub-lattices of $\mathcal{L}(\mathbf{B})$ are generated by, e.g, the following bases

$$\mathbf{B}_1 = \begin{bmatrix} 1 & -1 \\ -1 & 1 \end{bmatrix}, \quad \mathbf{B}_2 = \begin{bmatrix} 2 & 0 \\ 0 & 1 \end{bmatrix}, \quad \mathbf{B}_3 = \begin{bmatrix} 0 & 2 \\ 1 & 0 \end{bmatrix}. \tag{31}$$

In order to determine analytically a lower bound on the probability of converging to the correct solution, we need to prove the following lemma, which provides bounds on the number of sub-lattices.

*Lemma 5.4:* Given a lattice $\mathcal{L}_x$ of rank $N$ embedded in $\mathbb{R}^N$, the number $f_N(m)$ of sub-lattices of index $m$ is bounded by

$$m^{N-1} < f_N(m) < m^N. \tag{32}$$

*Proof:* It is possible to derive both an upper and a lower bound on the number of sub-lattices that are independent from the prime factorisation of $m$ starting from (30). Since for all cases of interest $N > 1$, we have:

$$\frac{p_i^{N+j-1} - 1}{p_i^j - 1} > \frac{p_i^{N+j-1}}{p_i^j}. \tag{33}$$

Substituting in (30), we have a function $\underline{f_N}(m)$ that is guaranteed to yield values below $f_N(m)$:

$$\underline{f_N}(m) = \prod_{i=1}^{q} \prod_{j=1}^{t_i} \frac{p_i^{N+j-1}}{p_i^j}. \tag{34}$$

This can be simplified to:

$$\underline{f_N}(m) = \prod_{i=1}^{q} p_i^{t_i(N-1)}. \tag{35}$$

This is equivalent to the $(N-1)^{\text{th}}$ power of the product of the prime factors of $m$. That is, the lower bound of $f_N(m)$ can be expressed as:

$$\underline{f_N}(m) = m^{N-1}. \tag{36}$$

In terms of the upper bound of $f_N(m)$, we proceed similarly by starting with the observation that:

$$\frac{p_i^{N+j-1} - 1}{p_i^j - 1} < \frac{p_i^{N+j}}{p_i^j}. \tag{37}$$

By substituting back into (30), we can observe that:

$$\prod_{i=1}^{q} \prod_{j=1}^{t_i} \frac{p_i^{N+j}}{p_i^j} = m \underline{f_N}(m). \tag{38}$$

Hence, it is easy to see that the upper bound on $f_N(m)$ can be expressed as:

$$\overline{f_N}(m) = m^N. \tag{39}$$

Therefore, since $\underline{f_N}(m) < f_N(m) < \overline{f_N}(m)$, we have:

$$m^{N-1} < f_N(m) < m^N. \tag{40}$$

∎

Now, consider a specific sub-lattice $\mathcal{L} \subset \mathcal{L}_x$ of index $m$ and a set of $P$ vectors from the original lattice $\mathcal{L}_x$. In the case of uniformly distributed vectors, the probability that one vector belong to the sub-lattice $\mathcal{L}$ is equal to $(1/m)$. Thus, the probability that all $P$ vectors belong to the same sub-lattice $\mathcal{L}$ is equal to $(1/m)^P$, assuming statistical independence among the set of vectors.

Let $p_{\text{fail}}(N, P)$ denote the probability of failing to detect the underlying lattice $\mathcal{L}_x$ of rank $N$, when $P$ points are observed. Then, $p_{\text{succ}}(N, P) = 1 - p_{\text{fail}}(N, P)$. A failure occurs whenever all $P$ vectors fall in any of the sub-lattices of index $m$. Hence, we can write

$$p_{\text{fail}}(N, P) < \sum_{m=2}^{\infty} f_N(m) \left(\frac{1}{m}\right)^P < \sum_{m=2}^{\infty} m^N \left(\frac{1}{m}\right)^P = \sum_{m=2}^{\infty} \frac{1}{m^{P-N}} = \zeta(P-N) - 1 \quad (41)$$

The first inequality is a union bound, i.e., the probability of failure is upper bounded by the sum of the probabilities of observing all $P$ vectors in a given sub-lattice. The second inequality follows from the upper bound given by Lemma 5.4. The last expression contains $\zeta(\cdot)$, which is the Riemann's zeta function. That is,

$$\zeta(s) = \sum_{m=1}^{\infty} \frac{1}{m^s}. \quad (42)$$

Note that the infinite series converges when the real part of the argument $s$ is greater than 1. In our case, this requires $P - N > 1$ or $P > N + 1$. Then, the probability of success is lower bounded by

$$p_{\text{succ}}(N, P) > 2 - \zeta(P - N). \quad (43)$$

It is interesting to observe that the probability of failure/success depend solely on the difference $P - N$. Hence, the number $P$ of observed vectors needed to correctly identify the underlying lattice grows linearly with the dimensionality $N$ of the embedding vector space, despite the number of potential lattice points grows exponentially with $N$, as indicated in Section IV.

Figure 3 shows that the upper bound on the probability of failure rapidly decreases to zero even for modest values of $P - N$.

## VI. EXPERIMENTS

Section V provided a lower bound on the probability of successfully identifying the transform and the quantization step sizes. In this section, this aspect is evaluated experimentally. In addition, we provide further insight on the complexity of the algorithm, expressed in terms of the number of recursive steps needed to compute the sought solution.

To this end, we generated data sets of $N$-dimensional vectors, whose elements are sampled from a Gaussian random variable $\mathcal{N}(0, \sigma^2)$. We considered the adverse case in which the elements are independent and identically distributed. Therefore, the distribution of the vectors is isotropic and no clue could be obtained from a statistical analysis of the distribution. Without loss of generality, we set $\sigma = 2$, $\mathbf{W} = \mathbf{I}$ and $\Delta_i = 1$, $i = 1, \ldots, N$. The same results were obtained using different transform matrices and quantization step sizes.

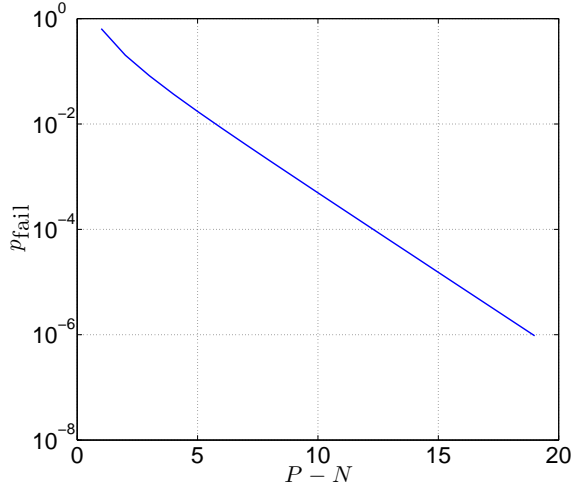

Fig. 3. Upper bound on the probability of failure $p_{\text{fail}}(P, N)$.

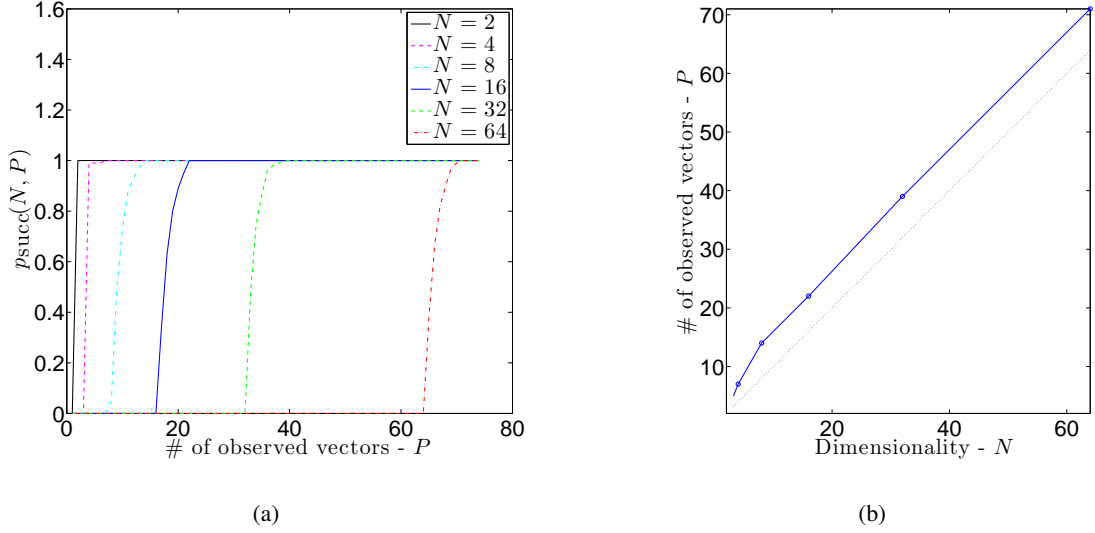

(a)            (b)

Fig. 4. (a) Empirical probability of success of Algorithm 1 in identifying the transform and the quantization step sizes as a function of the number of observed vectors $P$ and the dimensionality of the embedding vector space $N$. (b) Number of observed vectors $P$ needed to achieve $p_{\text{succ}}(N, P) > 1 - \epsilon$, with $\epsilon = 10^{-15}$.

Figure 4(a) shows the empirical probability of success when $N = 2, 4, 8, 16, 32, 64$, and the number of observed vectors $P$ is varied, averaged over 100 realizations. As expected $p_{\text{succ}}(N, P) = 0$ when the number of vectors $P$ does not exceed the dimensionality of the embedding vector space, i.e., $P \leq N$. Then, as soon as $P > N$, $p_{\text{succ}}(N, P)$ grows rapidly to one, when just a few additional vectors are visited. More specifically, Figure 4(b) illustrates the number of observed vectors $P$ needed to achieve $p_{\text{succ}}(N, P) > 1 - \epsilon$, where $\epsilon$ was set equal to $10^{-15}$. It is possible to observe that, when $N > 2$, the number of observed vectors needs to exceed by 6-7 units the dimensionality, and such a difference is independent from $N$, as expected based on the analysis in Section V. Note that the results shown in Figure 4 are completely oblivious of the specific implementation of Algorithm 2.

At the same time, it is interesting to evaluate the complexity when the specific instance of Algorithm 2, namely Algorithm 3, is adopted. Figure 5 shows the total number of recursive calls needed to converge to the solution of (15). Note that when a large enough number $P$ of vectors is observed, the algorithm converges to the correct lattice $\mathcal{L}_x$. Thus, visiting additional vectors does not increase the number of recursive calls, since the base step of the recursion is always met. Figure 5 shows two cases, that differ in the way the set of observed vectors is visited, i.e., randomly, or sorted in ascending order of distance from the origin of the vector space. In both cases, the number of recursive calls grows linearly with $N$. This is aligned with the analysis in Section V-B, which shows that convergence proceeds at a rate such that the number of recursive steps is upper bounded by $\lceil \log_2 |\mathcal{L}(\mathbf{B}^{(0)})|/|\mathcal{L}_x| \rceil$. A (loose) bound on the lattice determinant is given by

$$|\mathcal{L}(\mathbf{B}^{(0)})| = |\det(\mathbf{B}^{(0)})| \leq \|\mathbf{b}_1^{(0)}\|_2 \|\mathbf{b}_2^{(0)}\|_2 \cdot \|\mathbf{b}_N^{(0)}\|_2 \leq \|\mathbf{b}_{\max}^{(0)}\|_2^N, \tag{44}$$

where the first inequality stems from Hadamard inequality and $\mathbf{b}_{\max}^{(0)}$ is the column of $\mathbf{B}^{(0)}$ with the largest norm. Therefore,

$$\lceil \log_2 |\mathcal{L}(\mathbf{B}^{(0)})|/|\mathcal{L}_x| \rceil \leq \lceil N \log_2(\|\mathbf{b}_{\max}^{(0)}\|_2)/|\mathcal{L}_x| \rceil \tag{45}$$

This explain the dependency on $N$, as well as the fact that sorting the vectors so as to initialize $\mathbf{B}^{(0)}$ with shorter vectors reduces the number of recursive calls.

## VII. CONCLUSIONS

In this paper we proposed a method which is able to identify the parameters of a transform coder from a set of $P$ transform decoded vectors embedded in a $N$-dimensional space. We proved that it is possible to successfully identify the transform and the quantization step sizes when $P > N$ and this despite of the huge number of potential quantization bins, which grows exponentially with $N$ for a target bitrate. In addition, we proved that the probability of failure decreases exponentially to zero when $P - N$ increases. In our experiments we found that an excess of approximately 6-7 observed vectors beyond the dimension $N$ of the space is generally sufficient to ensure successful convergence.

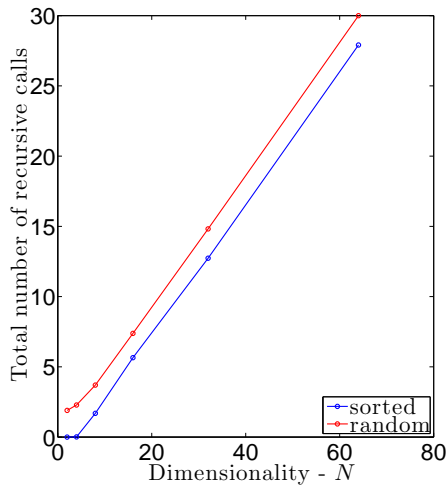

Fig. 5. Total number of recursive calls to `recurseTI` as a function of the dimensionality of the space $N$ and the strategy adopted to visit the observed vectors.

In this paper, we focused on a noiseless scenario, in which we observe directly the output of the decoder. In some cases, though, signals are processed in complex chains, in which multiple transform coders are cascaded, thus introducing noise in the observed vectors. Consequently, the observed vectors do not lie exactly on lattice points. Extending the proposed method to this new scenario represents an interesting research avenue to be investigated.

## REFERENCES

[1] G.K. Wallace, "The JPEG still picture compression standard," *IEEE Transactions on Consumer Electronics*, vol. 38, no. 1, pp. xviii–xxxiv, feb 1992.

[2] D.S. Taubman and M. W. Marcellin, *JPEG 2000: Image Compression Fundamentals, Standards, and Practice*, Springer-Verlag, 2002.

[3] D.T. Lee, "JPEG 2000: Retrospective and new developments," *Proceedings of the IEEE*, vol. 93, no. 1, pp. 32–41, jan 2005.

[4] G.J. Sullivan and T. Wiegand, "Video compression - from concepts to the H.264/AVC standard," *Proceedings of the IEEE*, vol. 93, no. 1, pp. 18–31, jan 2005.

[5] M. Winken, P. Helle, D. Marpe, H. Schwarz, and T. Wiegand, "Transform coding in the HEVC test model," in *IEEE International Conference on Image Processing (ICIP)*, 2011, pp. 3693–3696.

[6] J. Huang and P. Schultheiss, "Block quantization of correlated Gaussian random variables," *IEEE Transactions on Communications Systems*, vol. 11, no. 3, pp. 289–296, sep 1963.

[7] V. K. Goyal, "Theoretical foundations of transform coding," *IEEE Signal Processing Magazine*, vol. 18, no. 5, pp. 9–21, September 2001.

[8] M. Vetterli, "Wavelets, approximation, and compression," *IEEE Signal Processing Magazine*, vol. 18, no. 5, pp. 59–73, sep 2001.

[9] A. Cohen, I. Daubechies, O.G. Guleryuz, and M.T. Orchard, "On the importance of combining wavelet-based nonlinear approximation with coding strategies," *IEEE Transactions on Information Theory*, vol. 48, no. 7, pp. 1895–1921, jul 2002.

[10] S. Mallat, *A Wavelet Tour of Signal Processing*, Academic Press, 1998.

[11] V.K. Goyal, "Transform coding with integer-to-integer transforms," *IEEE Transactions on Information Theory*, vol. 46, no. 2, pp. 465–473, mar 2000.

[12] J. Lukas, J. Fridrich, and M. Goljan, "Digital camera identification from sensor pattern noise," *IEEE Transactions on Information Forensics and Security*, vol. 1, no. 2, pp. 205–214, June 2006.

[13] M. Chen, J. J. Fridrich, M. Goljan, and J. Lukás, "Determining image origin and integrity using sensor noise," *IEEE Transactions on Information Forensics and Security*, vol. 3, no. 1, pp. 74–90, 2008.

[14] T. Bianchi and A. Piva, "Image forgery localization via block-grained analysis of JPEG artifacts," *IEEE Transactions on Information Forensics and Security*, vol. 7, no. 3, pp. 1003–1017, June 2012.

[15] G. Valenzise, S. Magni, M. Tagliasacchi, and S. Tubaro, "No-reference pixel video quality monitoring of channel-induced distortion," *IEEE Transactions on Circuits and Systems for Video Technology*, vol. 22, no. 4, pp. 605–618, 2012.

[16] M. Naccari, M. Tagliasacchi, and S. Tubaro, "No-reference video quality monitoring for H.264/AVC coded video," *IEEE Transactions on Multimedia*, vol. 11, no. 5, pp. 932–946, 2009.

[17] M.R. Banham and A.K. Katsaggelos, "Digital image restoration," *IEEE Signal Processing Magazine*, vol. 14, no. 2, pp. 24–41, 1997.

[18] Z. Fan and R. L. de Queiroz, "Identification of bitmap compression history: JPEG detection and quantizer estimation," *IEEE Transactions on Image Processing*, vol. 12, no. 2, pp. 230–235, 2003.

[19] J. Lukás and J. Fridrich, "Estimation of primary quantization matrix in double compressed JPEG images," in *Proc. of DFRWS*, 2003.

[20] Y. Chen, K. S. Challapali, and M. Balakrishnan, "Extracting coding parameters from pre-coded MPEG-2 video," in *IEEE International Conference on Image Processing (ICIP)*, 1998, pp. 360–364.

[21] H. Li and S. Forchhammer, "MPEG2 video parameter and no reference PSNR estimation," in *Picture Coding Symposium (PCS)*, 2009, pp. 1–4.

[22] W. Luo, M. Wu, and J. Huang, "MPEG recompression detection based on block artifacts," in *Society of Photo-Optical Instrumentation Engineers (SPIE) Conference Series*, 2008, vol. 6819 of *Society of Photo-Optical Instrumentation Engineers (SPIE) Conference*.

[23] W. Wang and H. Farid, "Exposing digital forgeries in video by detecting double quantization," in *Proceedings of the 11th ACM workshop on Multimedia and security*, New York, NY, USA, 2009, MM&Sec, pp. 39–48, ACM.

[24] M. Tagliasacchi and S. Tubaro, "Blind estimation of the QP parameter in H.264/AVC decoded video," in *International Workshop on Image Analysis for Multimedia Interactive Services (WIAMIS), 2010*, 2010, pp. 1–4.

[25] P. Bestagini, A. Allam, S. Milani, M. Tagliasacchi, and S. Tubaro, "Video codec identification," in *IEEE International Conference on Acoustics, Speech, and Signal Processing (ICASSP)*, 2012, pp. 2257–2260.

[26] R. M. Gray and D.L. Neuhoff, "Quantization," *IEEE Transactions on Information Theory*, vol. 44, no. 6, pp. 2325–2383, oct 1998.

[27] R. Zamir, S. Shamai, and U. Erez, "Nested linear/lattice coes for structured multiterminal binning," *IEEE Transactions on Information Theory*, vol. 48, no. 6, pp. 1250–1276, june 2002.

[28] D. Wübben, D. Seethaler, J. Jaldén, and G. Matz, "Lattice reduction: A survey with applications in wireless communications," *IEEE Signal Processing Magazine*, vol. 28, no. 3, pp. 70–91, May 2011.

[29] R. Neelamani, R. de Queiroz, Z. Fan, S. Dash, and R.G. Baraniuk, "JPEG compression history estimation for color images," *IEEE Transactions on Image Processing*, vol. 15, no. 6, pp. 1365–1378, June 2006.

[30] A. K. Lenstra, H. W. Lenstra, and L. Lovász, "Factoring polynomials with rational coefficients," *Mathematische Annalen*, vol. 261, pp. 515–534, 1982, 10.1007/BF01457454.

[31] J. H. Conway, N. J. A. Sloane, and E. Bannai, *Sphere-packings, lattices, and groups*, Springer-Verlag New York, Inc., New York, NY, USA, 1987.

[32] H. Cohen, *A Course in Computational Algebraic Number Theory*, vol. 138 of *Graduate Texts in Mathematics*, Springer, 1993.

[33] B. Gruber, "Alternative formulae for the number of sublattices," *Acta Crystallographica Section A*, vol. 53, pp. 807–808, 1997.